# Intrusion Detection System with Machine Learning and Multiple Datasets

**Haiyan Xuan[1], Mohith Manohar[2*]**

[1]Carmel High School, Carmel, IN, United States of America

[2]SimplyFixable, New York, NY, United States of America

[2]Department of Computer Science, Columbia University, New York, NY, United States of America

**\* Correspondence:**
Mohith Manohar
mm5874@columbia.edu

**Abstract**

As Artificial Intelligence (AI) technologies continue to gain traction in the modern-day world, they ultimately pose an immediate threat to current cybersecurity systems via exploitative methods. Prompt engineering is a relatively new field that explores various prompt designs that can hijack large language models (LLMs). If used by an unethical attacker, it can enable an AI system to offer malicious insights and code to them. In this paper, an enhanced intrusion detection system (IDS) that utilizes machine learning (ML) and hyperparameter tuning is explored, which can improve a model's performance in terms of accuracy and efficacy. Ultimately, this improved system can be used to combat the attacks made by unethical hackers. A standard IDS is solely configured with pre-configured rules and patterns; however, with the utilization of machine learning, implicit and different patterns can be generated through the models' hyperparameter settings and parameters. In addition, the IDS will be equipped with multiple datasets so that the accuracy of the models improves. We evaluate the performance of multiple ML models and their respective hyperparameter settings through various metrics to compare their results to other models and past research work. The results of the proposed multi-dataset integration method yielded an accuracy score of 99.9% when equipped with the XGBoost[1] and random forest[2] classifiers and RandomizedSearchCV[3] hyperparameter technique.

## 1      Introduction

### 1.1      Problem Statement

The emergence of newly advanced LLMs poses a potential danger to existing IDS and other monitoring systems from unforeseen attacks from LLM-targeted exploitative methods. A new study has shown how injected prompts can alter the behavior of LLMs through exploitative methods, including remote code execution, unwanted intrusion, and more (Greshake et al., 2023). Therefore, to combat this danger, a intrusion detection system enhanced with machine learning is necessary to achieve evaluation metrics that perform highly.

---

[1] https://xgboost.readthedocs.io/en/stable/

[2] https://scikit-learn.org/stable/modules/generated/sklearn.ensemble.RandomForestClassifier.html

[3] https://scikit-learn.org/stable/modules/generated/sklearn.model_selection.RandomizedSearchCV.html

## 1.2 Network

A network is the accumulation of the interconnections between two or more systems. Networks can vary in size, depending on the number of devices connected to them as well as the use case for them. Smaller units of data, called packets, travel within networks between devices. Packets carry crucial information, such as a request to retrieve a webpage's content from a web server, a print job request, etc. In this research, we discuss these packets in the form of datasets that contain typical fields and information, like a packet.

## 1.3 IDS

In essence, an IDS monitors and detects malicious network activities. A standard IDS utilizes patterns and rules to detect whether a packet is malicious. It must be able to process network activities on a large scale and quickly. However, it is crucial to note that the IDS itself has no capability of preventing or resolving malicious activities. Therefore, an administrator is typically responsible for resolving the issue. There are two types of IDS: host-based (HIDS) and network-level (NIDS). In a host-based IDS, the system is deployed at one of the device's endpoints, allowing the system to monitor the specifics of the device, including its specific network traffic, running processes, etc. On the other hand, a network-level IDS is deployed onto the entire network, which means that it can review all the network traffic that flows through the network. NIDS, however, does not have the full capability of "knowing" every single detail on the host-based level.

## 1.4 ML with IDS

Machine learning is a prevalent field that essentially enables computers to learn patterns and parameters from a dataset, which can be used to make predictions based on those datasets. There are two main methods that machine learning can utilize: supervised learning and unsupervised learning. In supervised learning, there is a clear input and output, and the dataset is divided into testing and training datasets. On the other hand, unsupervised learning is mainly to explore and discover new patterns and trends in a dataset and is completely up to the model to develop the patterns. In this paper, supervised learning will be utilized to ensure that the chosen ML model is able to correctly identify if a packet is benign or malicious. Given the fields and entries of IPV4 packets, the model will be responsible for drawing the connections and patterns given the input and output. The ultimate purpose is to increase the accuracy score while also maintaining a sensible confusion matrix, which measures how well the model can correctly identify if a packet is malicious or benign.

## 1.5 Practical Aspects of ML

### 1.5.1 Preprocessing

Before the dataset is loaded into the model, there are a few steps that are required in terms of preprocessing: splitting, encoding, class imbalance handling, and feature selection. Splitting is the process of splitting the dataset into two sub-groups: training and testing. The training dataset is predominantly used to train the machine learning model. On the other hand, the testing dataset is used to measure the performance of the model. ML models require numerical values to process the dataset; therefore, encoding is required to convert non-numerical values, such as strings and empty values, into numerical values. Datasets may have a class imbalance where a majority and a minority class can exist. Therefore, class imbalance handling tackles this problem by generating synthetic instances in the minority class using a $k$-nearest neighbors algorithm, ultimately augmenting the minority class and balancing the dataset. Consequently, the performance of the machine learning models will be enhanced. Lastly, feature selection involves the process of choosing a subgroup of a dataset's features to effectively train an ML model.

### 1.5.2 Hyperparameter Tuning

Hyperparameter tuning is the process of searching for and identifying the best set of hyperparameter configurations (Mahfouz et al., 2020). Hyperparameters are external settings that are used to modify the behavior and efficacy of how the parameters are learned. Parameters are internal configurations that are set after a model has learned. In essence, parameters are deduced by the learning of the ML model, whereas hyperparameters are manually set by the programmer. Typically, hyperparameter tuning involves two main methods: manual or algorithmic. The manual method involves tweaking hyperparameter settings by hand, and it can be quite tedious to do. There are three main algorithms that can be used as an alternative to the manual approach: Bayesian optimization, grid search, and random search. The Bayesian optimization method is a more efficient method that identifies the best set of configurations based on previous runs, probability, and regression analysis. Grid search involves brute-forcing all combinations of hyperparameter settings to find the best set. Obviously, this implies that this method is more resource-intensive and time-consuming. Finally, a random search involves random configurations and returns the best set after the iterations.

### 1.6 Previous Work

There have been much past research works that have been conducted to devise various methods and techniques to produce an IDS with evaluation metrics that perform highly. With the growing prevalence of AI, many methods adopt AI-based principles to create an accurate IDS, and some utilize hyperparameter tuning techniques. Wazirali (2020) performed k-nearest neighbor hyperparameter tuning on the NSL-KDD dataset, which yielded an accuracy score of 99.1%. Dhaliwal et al. (2018) discovered that the XGBoost library on the NSL-KDD dataset worked better than other classifiers. Kanimozhi and Jacob (2019) used artificial neural networks and hyperparameter tuning, which resulted in a high accuracy score of 99.97% with the GridSearchCV technique.

Halimaa A. and Sundarakantham (2019) explored two main machine learning techniques, support vector machines (SVM) and naive bayes, and determined that SVM worked better than naive bayes. Patgiri et al. (2018) observed that random forest classifier performed better than the SVM classifier before feature selection. Ugochukwu and Bennett (2018) also determined that random forest and random tree algorithms generally perform better on the KDDCup99 dataset than Bayes Net and J48 algorithms.

### 1.7 Contributions

In this paper, we propose a method that involves equipping multiple datasets during the training and testing stages to increase the accuracy and performance of the chosen ML models. We apply hyperparameter tuning to each of the ML models to enhance their performance. We then evaluate the performance of the model through various metrics and compare the results.

### 2 Method

## 2.1 Dataset Description

One of the datasets that is utilized in this research is the UNSW-NB15 dataset, which was collected by Cyber Range Lab of UNSW Canberra (Moustafa and Slay, 2015). In this research, the 1st out of 4 CSV files has been used. The dataset features real-world and synthetic data that contains more than 600,000 malicious/benign, real-life network packets, with 3% of the dataset containing malicious packets. It has 49 columns that detail various attributes of each packet, some of which pertain to each packet, and some of which classify the packet. The columns include basic attributes, such as source/destination IP addresses, time-to-live (TTL) values, record total duration, etc. The UNSW-NB15 is also a multi-class dataset that contains detailed information regarding which category a malicious packet falls under (Fuzzers, Analysis, Backdoors, DoS Exploits, Generic, Reconnaissance, Shellcode, and Worms). Sydney (2015) provides a thorough list of the features.

Another dataset that is utilized is the KDDCup 1999 dataset. The KDDCup 1999 dataset contains around 4,900,000 entries, comprising both benign and malicious packets. It is a popular dataset that has been used extensively to evaluate the effectiveness of anomaly-based detection systems (Tavallaee et al., 2009). The dataset contains 41 features that describe each capture network packet, including duration, service type, etc. Lab (1999) provides the full and detailed list of the KDD-CUP 1999 dataset.

## 2.2 Hardware Specifications

The code was executed on Google Colab using Python 3 Runtime Type, CPU Hardware Accelerator, 51 GB of RAM, and 225.8 GB of disk storage.

## 2.3 Dataset Preprocessing

By default, the dataset itself isn't prepared to be deployed into the ML models. One of the issues is the presence of non-numerical values, which include dashes (in replacement of null values), empty values, words, etc. To combat this, we apply encoding to the dataset to convert non-numerical values to numerical values by iterating through the entries and using the appropriate conversion method. Afterwards, we divide the dataset into training and testing datasets. Lastly, the chosen datasets have a class imbalance, meaning that a small percentage of the dataset contains malicious packets. **Figure 1** illustrate the distribution of the number of benign and malicious packets in the UNSW-NB15 and KDDCup 1999 datasets. **Figures 1a and 1c** show the existence of the class imbalance, therefore we select and utilize the SMOTE[4] library to create a more balanced dataset of malicious and benign packets. **Figure 1b** indicates the increased number of malicious packets to rebalance the dataset, whereas **Figure 1d** depicts that the benign class increased in quantity to rebalance the dataset. To combine the two datasets together, both datasets must have the same number of features; therefore, we utilize feature selection to choose the most relevant, similar datasets to concatenate the two models together and train the chosen ML models. The full list of the chosen features can be found in **Table 1**.

---

[4] https://imbalanced-learn.org/dev/references/generated/imblearn.over_sampling.SMOTE.html

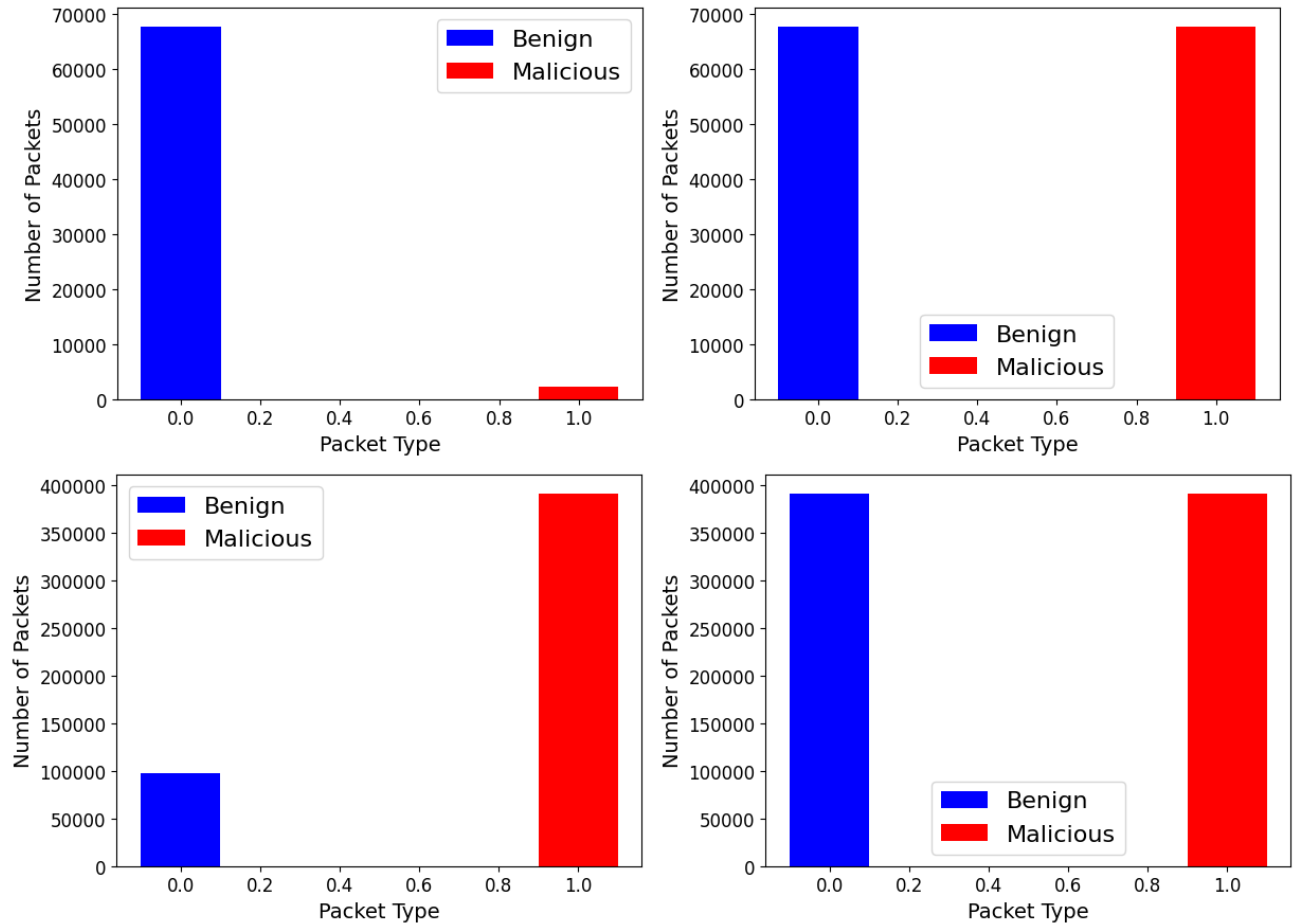

**Figure 1** **(A)** Bar graph that depicts the distribution of benign (blue) and malicious (red) packets before applying SMOTE library to the UNSW-NB15 dataset. **(B)** Bar graph that depicts the distribution of benign (blue) and malicious (red) packets after applying SMOTE library to the UNSW-NB15 dataset. **(C)** Bar graph that depicts the distribution of benign (blue) and malicious (red) packets before applying SMOTE library to the KDD-CUP 1999 dataset. **(D)** Bar graph that depicts the distribution of benign (blue) and malicious (red) packets after applying SMOTE library to the KDD-CUP 1999 dataset.

### 2.4 ML Models

The following ML models are considered for this research work:

1. Gradient Boosting Machines (XGBoost)
2. Logistic Regression[5]
3. Random Forest Classifier

### 2.5 Hyperparameter Tuning

We equip the RandomizedSearchCV technique for all the ML models to efficiently determine the best hyperparameter setting. The optimal hyperparameter settings for the XGBoost is as follows: n_estimators are 200, max_depth is 3 and learning_rate is 0.1. For logistic regression, when penalty is L2 and when C is 100, the performance is optimal. When the n_estimators is 200 and min_samples_leaf is 4, the random forest classifier achieves its best performance.

---

[5] https://scikit-learn.org/stable/modules/generated/sklearn.linear_model.LogisticRegression.html

## 3 Results

### 3.1 Evaluation Metrics

To measure the performance of each model, we record the values in **Table 2**, with the training set consisting of 10% of the total data. Additionally, the different evaluation metrics were recorded with varying training dataset sizes (2%, 4%, 6%, 8%, and 10%) to observe any key patterns and observations. We record each of these metrics for the ML models in **Figure 2**. We acknowledge that usually the training set is larger than the testing set, but in our case, we utilized a smaller training set. The following evaluation metrics are recorded for each model, and the equation for them are shown below.

$$1. Accuracy = \frac{Number\ of\ Correct\ Predictions}{Total\ Number\ of\ Predictions}$$

$$2. Precision = \frac{True\ Positives}{True\ Positives + False\ Positives}$$

$$3. Recall = \frac{True\ Positives}{True\ Positives + False\ Negatives}$$

$$4. F1\ Score = \frac{2 * Precision * Recall}{Precision + Recall}$$

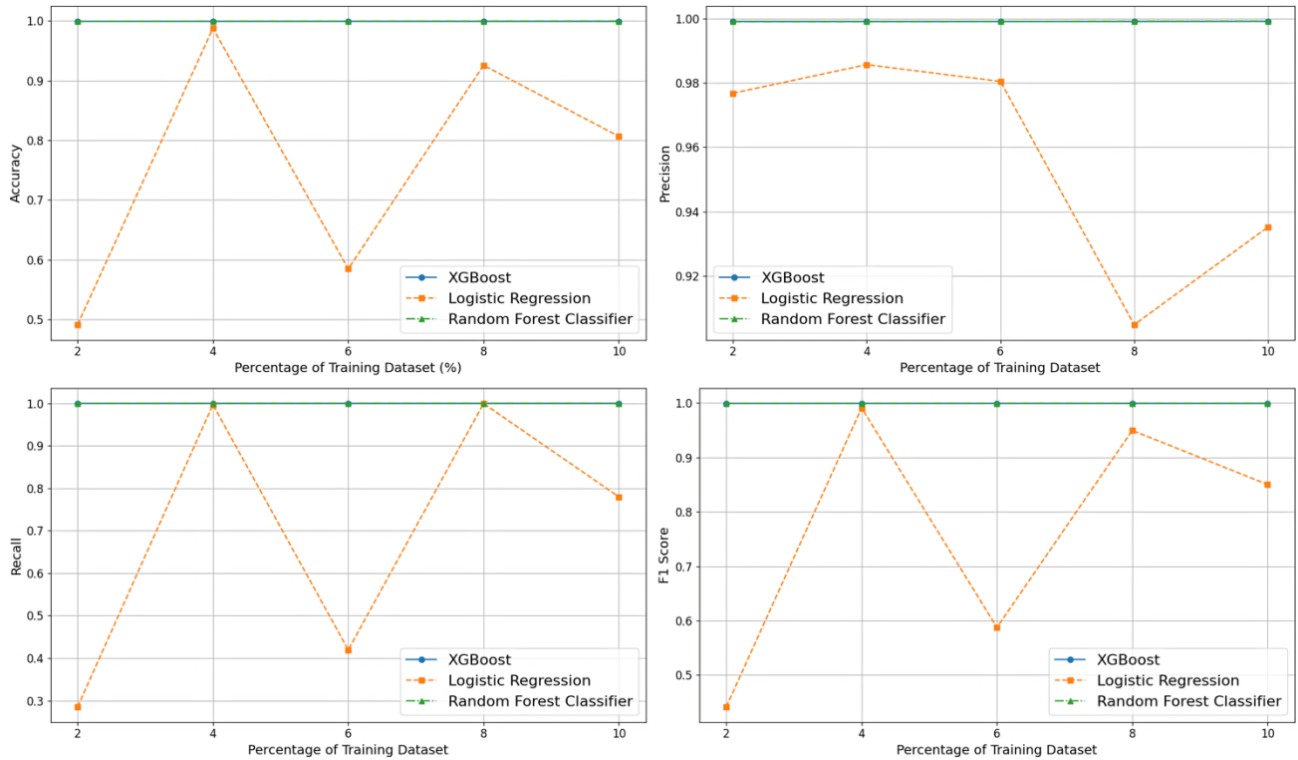

**Figure 2 (A)** Line graph that features the accuracy score of the progressive learning for each of the selected ML models. **(B)** Line graph that features the precision score of the progressive learning for each of the selected ML models. **(C)** Line graph that features the recall score of the progressive learning for each of the selected ML models. **(D)** Line graph that features the F1 score of the progressive learning for each of the selected ML models.

### 3.2 Overall Performance

It can be clearly seen from the evaluation metrics in **Table 2** that XGBoost and random forest performs more accurately compared to the logistic regression model when using the RandomizedSearchCV method. Additionally, **Figure 2** supports the observation that the XGBoost and random forest models perform exceptionally well for any of the tested and varying training dataset, whereas the performance of the logistic regression varies depending on the size of the training dataset. From **Table 2**, the XGBoost and random forest classifiers had an accuracy score of was 99.9%, and the logistic regression model had an accuracy score of 80.6%. Therefore, our proposed method of combining multiple datasets together yields a better accuracy score.

## 4 Discussion

### 4.1 Conclusion

Certain models perform better than other models. Specifically, the XGBoost and random forest classifiers perform better than the logistic regression model. The accuracy score from the combined dataset approach results in 99.91% and 99.93% for the XGBoost and random forest classifiers, respectively. The accuracy of the IDS, particularly the XGBoost model, has improved with multiple datasets and hyperparameter tuning using the RandomizedSearchCV technique. XGBoost trained only on NSL-KDD dataset has an accuracy score of around 98.7% (Dhaliwal et al., 2018).

## 4.2 Future Work

To truly see if the performance of the model is sufficient to combat against AI-generated network activities, further research must be done to see if the discussed models are effective against abnormal network attacks as well. Research on prompt engineering attacks is limited, since LLMs were introduced to the public relatively recently; therefore, further investigation is required to investigate how LLMs can be exploited, so that these abnormal techniques can be considered for further improvements to the IDS. Additionally, various hyperparameter techniques can be used and compared to identify the best technique for each of the chosen models in this research.

**Table 1 List of the selected features used to combine the UNSW-NB15 and KDD-CUP 1999 datasets.**

| UNSW-NB15 | KDD-CUP 1999 |
| --- | --- |
| Dintpkt | duration |
| Djit | src_bytes |
| Dload | dst_bytes |
| Dpkts | land |
| Sintpkt | wrong_fragment |
| Sjit | urgent |
| Sload | hot |
| Spkts | num_failed_logins |
| Stime | logged_in |
| ackdat | num_compromised |
| ct_dst_ltm | root_shell |
| ct_dst_sport_ltm | su_attempted |
| ct_dst_src_ltm | num_root |
| ct_src_dport_ltm | num_file_creations |
| ct_src_ltm | num_shells |
| ct_srv_dst | num_access_files |
| ct_srv_src | num_outbound_cmds |
| ct_state_ttl | is_host_login |
| dbytes | is_guest_login |
| dloss | count |
| dmeansz | srv_count |
| dsport | serror_rate |
| dstip | srv_serror_rate |
| dttl | rerror_rate |
| dur | srv_rerror_rate |
| dwin | same_srv_rate |
| is_ftp_login | diff_srv_rate |
| is_sm_ips_ports | srv_diff_host_rate |
| proto_icmp | dst_host_count |
| proto_tcp | dst_host_srv_count |
| proto_udp | dst_host_same_srv_rate |
| res_bdy_len | dst_host_diff_srv_rate |
| sbytes | dst_host_same_src_port_rate |
| service_ftp | dst_host_srv_diff_host_rate |
| service_ftp-data | dst_host_serror_rate |
| service_http | dst_host_srv_serror_rate |
| service_irc | dst_host_rerror_rate |
| service_pop3 | dst_host_srv_rerror_rate |
| service_smtp | protocol_type_icmp |
| service_ssh | protocol_type_tcp |
| sloss | protocol_type_udp |
| smeansz | service_IRC |
| sport | service_ftp |
| srcip | service_ftp_data |
| sttl | service_http |
| swin | service_pop_3 |
| synack | service_smtp |
| tcprtt | service_ssh |

**Table 2 Evaluation metrics of the chosen ML models using the combined dataset with 10% size training dataset.**

| | Accuracy | Precision | Recall | F1 Score |
|---|---|---|---|---|
| XGBoost | 0.9991 | 0.9991 | 0.9996 | 0.9994 |
| Logistic Regression | 0.8064 | 0.9351 | 0.7796 | 0.8503 |
| Random Forest | 0.9993 | 0.9992 | 0.9998 | 0.9995 |